%% file: forArxiv.tex
\documentclass[aps,reprint,pra,superscriptaddress]{revtex4-2}

\usepackage[utf8]{inputenc}
\usepackage{graphicx}
\usepackage{amssymb}
\usepackage[svgnames,table]{xcolor}
\usepackage{rotating}
\usepackage{fancyhdr}
\usepackage{graphicx}
\graphicspath{{figures/}}
\usepackage[caption=false]{subfig}
\usepackage{booktabs}
\usepackage{natbib}
\usepackage{mathtools}
\usepackage[USenglish]{babel}
\usepackage[autostyle,autopunct=true]{csquotes}
\usepackage{siunitx}
\usepackage{hyperref}
\hypersetup{
    pdftoolbar=false,        
    pdfmenubar=true,        
    pdffitwindow=true,     
    pdfstartview={FitH},    
    pdfnewwindow=true,      
    colorlinks=true,       
    linkcolor=SlateBlue,          
    citecolor=LightSlateBlue,        
    filecolor=magenta,      
    urlcolor=Navy           
}

\DeclareSIPostPower \tothefourth{4}

\MakeOuterQuote{"}
\DeclarePairedDelimiter\ket{\lvert}{\rangle}

\newcommand\numberthis{\addtocounter{equation}{1}\tag{\theequation}}

\begin{document}
\title{Optical Refrigeration for an Optomechanical Amplifier}
\author{Samuel Schulz}
\affiliation{Department of Physics and Astronomy, Amherst College, Amherst, Massachusetts 01002}
\author{Yehonathan Drori}
\author{Christopher Wipf}
\author{Rana X Adhikari}
\affiliation{LIGO Laboratory, California Institute of Technology, Pasadena, California 91125}
\date{\today}
\begin{abstract}
We explore the viability of using optical refrigeration as a low-vibration cooling method for a phase-sensitive optomechanical amplifier proposed to improve the sensitivity of future gravitational wave detectors. We find that with moderate improvements on coolants currently available, optical refrigeration can improve the amplifier gain by a factor of 10 relative to what is possible with radiative cooling. We also show that the technique does not add significant noise to the amplifier. These results indicate that optical refrigeration can play an important role in cooling optomechanical devices.
\end{abstract}
\maketitle

\section{Introduction}
\subsection{Motivation}
Since the first direct detection of gravitational waves (GW) in 2015~\cite{GW150914}, the LIGO and Virgo interferometric GW detectors~\cite{theligoscientificcollaborationAdvancedLIGO2015, Acernese_2014} have amassed a catalog of 90 candidate GW sources~\cite{gwtc-3} across three observing runs. Future detectors, such as Voyager, NEMO, The Einstein Telescope, and possibly Cosmic Explorer~\cite{VoyagerInst:2020, reitzeCosmicExplorerContribution2019}, will use cryogenic silicon mirrors to extend the reach of observations, which is currently limited by thermal and quantum noise in the instruments. In these new detectors, quantum noise is expected to be the dominating source of noise throughout the detection band.


One way to mitigate quantum noise is by increasing the laser power circulating in each of the interferometer arms. Because Voyager aims to operate at cryogenic temperatures, any increase in cavity laser power must have a corresponding increase in cooling power of the test masses in order to combat heating caused by absorption.
Another way is using squeezed states of light injected into the interferometer's dark port~\cite{O3:Squeezing:2019}. However, the effectiveness of any quantum noise reduction is still limited by losses in interferometer optics~\cite{miao2019quantum}. To combat this, a phase-sensitive optomechanical amplifier (PSOMA) has been proposed, capable of mitigating optical losses in the readout and reducing quantum noise~\cite{bai2020phase}. Like the rest of the interferometer, the optics used for PSOMA must be kept near \SI{123}{\kelvin} because the device will use the same silicon mirrors as the rest of Voyager.



Clearly, both of these ways of reducing quantum noise require cooling. Because any introduced vibration to a suspended optic would increase noise, cryocooling techniques must introduce little to no vibration to the system. Vibrations are harmful even if none of the vibrations is transferred to a suspended optic. This is because some laser light might be scattered off the test mass onto surrounding objects before recombining with the main beam, thus introducing noise. If those objects are vibrating, a phase shift could be introduced along with fluctuations in radiation pressure, both of which would contribute to the noise~\cite{shapiro_cryogenically_2017}. Both of these conditions put harsh constraints on what types of cryocoolers are usable for LIGO-suspended optics. Although thermoelectric coolers (TECs) are vibration-free, existing TECs do not cool at cryogenic temperatures, and optical refrigeration has the potential for more cooling power than proposed TECs at low temperature~\cite{seletskiy_laser_2016}. Radiative cooling can often be useful, but we show here its limits for cooling small loads, which are essential to the optomechanical function of PSOMA. 

One potential solution to this problem is to use a solid-state optical refrigerator, which is nearly vibration-free and can cool at cryogenic temperatures, to cool parts of the interferometer. Recently an all-solid-state optical refrigerator was used to cool a sensor to 135~K from room temperature. Here we present the design consideration for an optical refrigerator for an optomechanical amplifier and show that it can outperform radiative cooling by a factor of 10 using near-future technology and that the noise introduced by such a technique is negligible. It should be noted that this analysis is not limited to PSOMA and can be readily applied in general to cryogenic optomechanical devices.

\subsection{Optical Refrigeration}

Solid-state optical refrigeration (OR) uses anti-Stokes fluorescence to cool material with applied optical power. Optical refrigeration has been demonstrated in several rare-earth-doped crystals~\cite{seletskiy_laser_2016}. A review of the underlying processes and OR materials can be found in Ref.~\cite{seletskiy_laser_2016}. 

\begin{figure}[h]
	\begin{center}
	\includegraphics[width=\columnwidth]{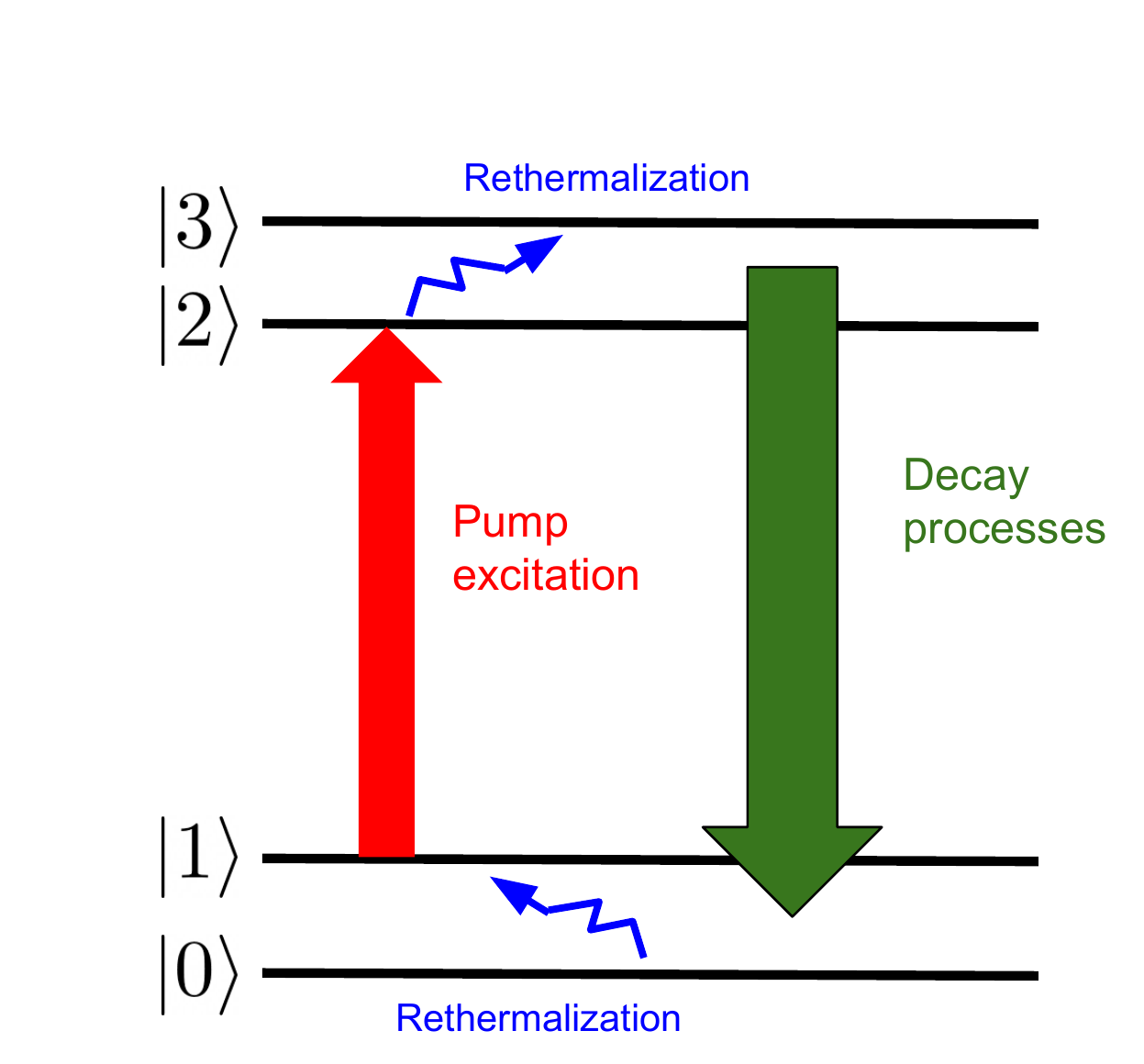}
	\caption{A simplified four-level model showing the cooling cycle for optical refrigeration. Dopant ions initially arrange themselves in a Boltzmann distribution with some atoms in $\ket{1}$. These are optically excited to $\ket{2}$ then some are excited via phonon absorption to $\ket{3}$ before decay to $\ket{0}$ and $\ket{1}$, restarting the cycle.}
	\label{fig:4LvlModel}
	\end{center}
	\end{figure}

The cooling power provided by a crystal in an optical refrigerator is given by
\begin{equation}
P_{cool}= \eta_c P_{crys},
\end{equation}
where $P_{crys}$ is the power absorbed by the crystal and
\begin{equation*}
\eta_c=\eta_{abs} \eta_e \frac{h \nu_f}{h \nu} -1,
\end{equation*}
is the cooling efficiency for external quantum efficiency $\eta_e$, absorption efficiency $\eta_{abs}$, pump frequency $\nu$ and mean fluorescence frequency $\nu_f$ where $h$ is the Planck constant~\cite{seletskiy_laser_2016}. 

The material properties which affect the cooling power for a given crystal are the external quantum efficiency, absorption efficiency (and its constituent parts background absorption $\alpha_b$ and dopant absorption $\alpha$), and mean fluorescence frequency, all of which, in general, are functions of crystal temperature.

Cooling power is also affected by the absorption of stray fluorescence, which would be Stokes-shifted on average and heat the crystal. This effect is explored in Appendix~\ref{sec:reabsorption}. We find that the additional heating load due to the absorption of stray fluorescence is negligible in the cases described here.

\section{Applicability for LIGO}

\subsection{Radiative Cooling Limits}
We first show the limitations of an existing alternative to OR, radiative cooling. Systems like PSOMA employ small mirrors to increase optomechanical coupling, which puts a limit on the degree to which they can be cooled radiatively. To find that limit in the case of PSOMA, we use the Stefan-Boltzmann law to find the radiated power $P$ for each mirror surface,
\begin{equation}
P=A \sigma \varepsilon T^4
\end{equation}
where $A$ is the area of that surface, $\varepsilon$ is the emissivity of that surface, $\sigma$ is the Stefan--Boltzmann constant and $T$ is the temperature at the surface, which for PSOMA will be \SI{123}{\kelvin}. If we assume that the mirror surface will be circular with a diameter about 3 times the beam diameter of \SI{10}{\milli \meter} and that the mass of the mirror will be $\SI{30}{\gram}$, as specified in Ref.~\cite{bai2020phase}, we can find the exact dimensions of a cylindrical mirror with those properties. Using $\varepsilon$ values of 0.5 and 0.9 for the dielectric coating and silicon surfaces, we can find the power radiated by each surface and sum them to find the total radiated power. These reflectivity numbers are those used for the LIGO test masses in~\cite{VoyagerInst:2020} and represent ideal or near-ideal figures. This means that our estimate for the maximum radiated power is likely an overestimate. A simple calculation finds a radiated power of $\sim\SI{40}{\milli\watt}$ at \SI{123}{\kelvin}. Assuming 1\,ppm absorption by the mirror, this would allow for a power circulating in the cavity of up to $\sim\SI{40}{\kilo\watt}$. For a $\SI{10}{\gram}$ mirror with the same properties otherwise, a maximum of $\sim\SI{19}{\milli\watt}$ could be cooled radiatively, allowing for a circulating power of $\sim\SI{19}{\kilo\watt}$. It could be possible to surpass this radiative cooling limit by using cooling fins, which increase the effective surface area of the mirror. Designs for cooling fins are in the early stages, and it is not yet clear how large of an effect they might have on the radiative cooling limit for small optomechanics.
\subsection{Optical Refrigeration}

For cooling silicon mirrors like those in PSOMA, the clear choice is to use a holmium-doped crystal because the silicon mirrors will have very low absorption across the fluorescence band of Ho$^{3+}$ ($\sim \SI{2}{\micro\meter}$). The minimum achievable temperature for any holmium-doped crystals that appear in currently published papers is not low enough to cool at \SI{123}{\kelvin}~\cite{rostami_observation_2019, rostami_observation_2021}. However, it is expected that upcoming holmium doped crystals will be able to cool at 123 K and below~\cite{rostami_observation_2019,rostami_observation_2021}. In particular, a recently investigated holmium-doped BYF crystal could cool to 80~K and below \cite{rostami_observation_2021} with improvements to its background absorption. Thus, for the calculations of cooling power here, we use improved figures for background absorption mentioned as reasonable improvements in Ref.~\cite{rostami_observation_2021}.

\begin{figure}[htp] 
\centering
\includegraphics[width=\columnwidth]{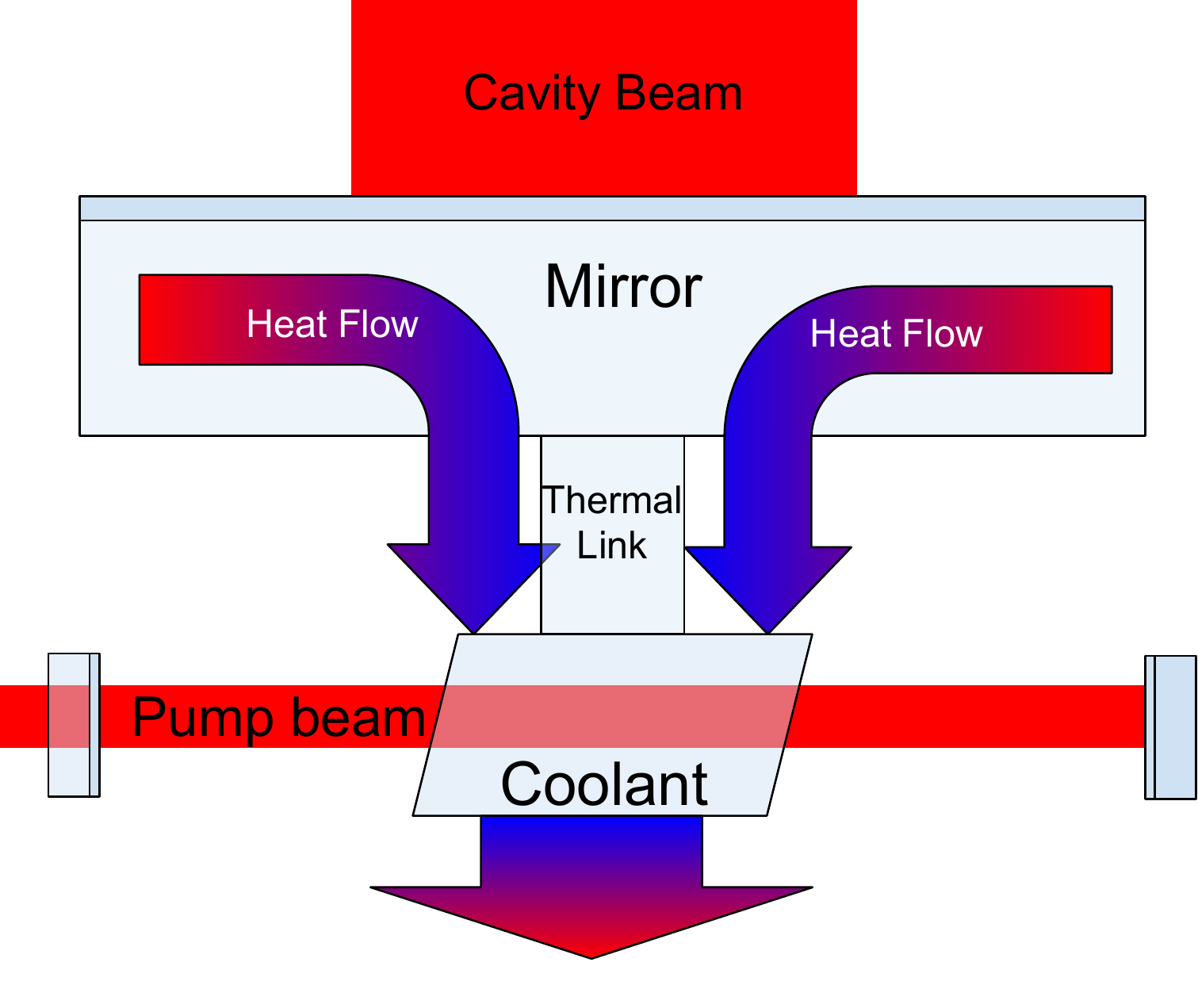}
\caption{A schematic for the proposed arrangement of using an optical coolant to cool PSOMA. An optional thermal link is included here. Even though the arrow representing the fluorescence is only pointed away from the mirror, the fluorescence, in reality, would be isotropic from all points in the path of the beam in the coolant.}
\label{fig:PSOMASchem}
\end{figure}

The first all-solid-state optical cryocooler used a thermal link made from YLF to shield the load~\cite{hehlen_first_2018}. Because the load will be largely transparent to the crystal's fluorescence, there is no inherent need to use a thermal link or shield the load, so the crystal can be adhered directly to the mirror with the optical path of the pump laser running perpendicular to the normal of the mirror surface, as shown in Figure \ref{fig:PSOMASchem}. However, a thermal link would reduce detrimental effects such as the absorption of stray fluorescence and induced noise, as will be further discussed in Section~\ref{results}. In our case, we could use a silicon thermal link that is optically contacted to the mirror because silicon is transparent to the crystal's fluorescence. For the calculations here, we assume a silicon thermal link with no bend to allow for an analytical understanding of the heat flow.


\subsection{Noise Considerations}
\label{sec:noiseconsider}
Attaching a coolant crystal to the back of the mirror could contribute noise in a few ways via the high-power light absorbed and emitted by the crystal.

\subsubsection{Absorption Imbalance Noise}
The first source of noise is noise in absorption events in the crystal. There will be a contribution of some classical noise due to the fact that absorption will be slightly higher (by about 0.5\,--\,0.9\% of the total absorbed power) on the odd-numbered passes through the crystal than the even ones and all of the odd passes will be on in one direction with the even ones in the other. This leaves a net force in one direction, which will have noise following the relative intensity noise of the pump laser, which we borrowed from existing calculations~\cite{bai2020phase}. Additionally, there is a shot noise due to the random nature of these absorption events, which we have included under absorption noise. Because the beam path for the OR pump beam is orthogonal to that of the cavity beam, displacement noise caused by absorption events would be orthogonal to the axis on which distance has to be precisely measured, so its effect would be greatly reduced in practice. We evaluated this noise as if it were in the direction along the cavity axis and will show that even in that case, it would not be enough to care about (see Sec.~\ref{results}).

\subsubsection{Fluorescence Recoil Noise}

The next source of the noise is the recoil due to the random fluctuations in the number of fluorescence events in the crystal. Because the average number of emitted photons heading away from the PSOMA mirror would be the same as the average number of emitted photons heading toward the PSOMA mirror, the only noise in the force due to these events would be quantum in nature. Any classical noise would contribute equally to both sides, so it can be ignored. This means that the amplitude spectral density (ASD) of the noise in the force due to recoil noise in the direction normal to the mirror surface path can be expressed as
\begin{equation}
F_r = \frac{2}{\pi}\frac{E_{photon}}{c} \sqrt{n_{photon}},
\end{equation}
where $E_{photon}$ is the average energy of an emitted photon, $c$ is the speed of light, $n_{photon}$ is the rate of emitted photons in units of \SI{}{\per\second}, and the factor of $2/\pi$ is the average value of $cos(\theta)$ where $\theta$ is the angle between the optical path of the emitted photon and the normal to the mirror surface. This force can be turned into a displacement using the mechanical transfer function of the mirror.
 
\subsubsection{Radiation Pressure Noise}
 
 Radiation pressure noise comes from fluctuations in the light which reflects off the mirror surface. Radiation pressure noise has already been explored for various LIGO applications, including PSOMA~\cite{bai2020phase,yu_quantum_2020,wang_boosting_2022,miao_enhancing_2015,cripe_measurement_2019}. This is the first time, however, that the radiation pressure noise caused by an optical refrigerator has been estimated. We have identified three potential sources of radiation pressure noise, explained below.
 
The next two sources are displacement noise sources due to the fluctuations in the fluorescence power reflecting off the reflective coating on the PSOMA mirror. Because the fluorescence of the crystal would be in the same band as that of the pump beam, we expect that almost all of the light will reflect off this surface. It should be noted that the use of a thermal link, which is allowed within our power budget, would significantly reduce noise due to light hitting the mirror surface. This is because a thermal link allows for additional shielding of fluorescence. Another way to limit this power could be by putting the crystal below the mirror so that much of the fluorescence misses the mirror. In this case, the light that hits the mirror would do so at a sharp angle, meaning most of the force would be upwards rather than along the cavity beam path. Therefore, the numbers presented in this report are likely pessimistic compared to what is possible with careful design choices which are beyond the scope of this work.
 
 To begin to evaluate the noise in this power, we must first get an idea of the average power hitting the mirror. To do so, we approximate the coolant crystal as a cube to find an approximation of the solid angle subtended by the mirror on the coolant crystal which then gives us the approximate proportion of the fluorescence power hitting the mirror surface using the well known differential solid angle formula. 
We integrated this over the mirror's surface, then averaged over the approximated volume of a cube. An additional prefactor of around 2 or 3 is also included to account for refraction at the coolant crystal-silicon boundary, which may increase the solid angle due to the relatively high index of refraction of silicon.

There are two places where noise could be introduced in this power: (1) fluctuations in the coolant pump laser intensity which would change the rates of absorption and therefore emission, and (2) "wave interaction noise" due to interference between light of slightly different frequencies~\cite{hodara_statistics_1965}. We will refer to (1) as relative intensity noise (RIN) and (2) as wave interaction noise (WIN). RIN is well studied and has been examined already for the cavity pump noise in PSOMA, so we will mirror those methods here using the power hitting the mirror surface as the relevant power~\cite{bai2020phase}. We found an expression for the displacement ASD due to wave interaction noise (See Appendix \ref{sec:WIN_noise}).
\begin{equation}
	A_{xx}\left(f\right) = \frac{1}{4\pi^2mcf^2}\sqrt{\frac{A_c}{A}}\frac{I_{tot}}{\sqrt{BW}},
\end{equation}
where $m$ is the mass of the mirror, $A_c$ is the coherence area of the fluorescence at the mirror surface, $A$ is the area of the mirror surface, $I_{tot}$ is the total power hitting the mirror surface, and $BW$ is the bandwidth of the fluorescence spectrum. 

The average coherence area is estimated using the approximation given in Eq. 2.4 of Ref.~\cite{mandel1965coherence}. This approximation is given by $A_c \sim \alpha (\frac{\lambda R}{\Delta l})^2$, where $\lambda$ is the mean fluorescence wavelength, $R$ is the distance from source to detector, $w$ is a defining width or radius of the source, and $\alpha \sim 1$ is a prefactor of order unity related to the geometry of the source. The approximation is meant to apply to the far field, which strictly does not describe the relationship between the mirror surface and the coolant crystal. However, because coherence area increases monotonically with distance from the source and is 0 at the source, this approximation will be an overestimate in the near field, giving us an upper limit on the WIN. Table \ref{tab:WIN_params} shows our estimated values for some relevant parameters. This fluctuation in the photon arrival count is then used to find the magnitude of the force, which is converted to a displacement noise magnitude. 

\begin{table}[h!]
    \centering
    \begin{tabular}{lr}
    \toprule
    Parameter  &  Estimated Value \\ \midrule
    Coherence area ($A_c$)    &  \SI{5.5}{\micro \meter \squared} \\
    Area of Mirror Surface ($A$) & \SI{7}{\centi \meter \squared} \\
    Emitter Linewidth ($BW$) & \SI{8.1}{\tera \hertz} \\
    \bottomrule
    \end{tabular}
\label{tab:WIN_params}
\caption{The estimated parameters used to calculate the wave interaction noise (WIN).}
\end{table}

To find the strain noises associated with these mirror displacement noises, we followed the approach outlined in Sec. III-C of Ref.~\cite{bai2020phase}.

\subsubsection{Leaked Light Noise}

There are a couple of noise sources that are not due to radiation pressure but rather related to leaked fluorescence into the readout chain, which might be confused for strain signal. First, some of the fluorescence which contacts the reflective surface of the mirror would couple to cavity modes in the PSOMA cavity. We provide a quantitative description of that noise in Section \ref{sec:leaked_light}. Second, scattered fluorescence light by surrounding surfaces might be scattered back into the PSOMA cavities, leading to scattered light noise. Baffling \ can largely mitigate this issue \cite{Kip:Scatter95}, especially when a thermal link is used, though the design details of a baffle are beyond the scope of this work.

\section{Results}
\label{results}

\subsection{Cooling Power}

We use existing spectroscopy, background absorption, and external quantum efficiency measurements as a basis for these calculations. The crystal we used for these calculations was the one detailed in Ref.~\cite{rostami_observation_2021}, with an external quantum efficiency of 0.997. We also used existing absorption data taken at 20~K intervals for that crystal~\cite{rostami_low-temperature_2021}. The effects of absorption saturation and amplified spontaneous emission, though theoretically investigated recently, are ignored here \cite{sheik-bahae_role_2022}. Additionally, because existing Ho:BYF crystals do not cool at cryogenic temperatures, we use predicted future background absorption values found in Ref.~\cite{rostami_observation_2021}. We provide results at 100~K to allow for the use of a thermal link, which would reduce various noise sources, as previously mentioned.

We find that the highest cooling performance at \SI{100}{\kelvin} would be when the crystal is pumped at $\sim \SI{2065}{\nano\meter}$. Lasers operating at $\SI{2070}{\nano\meter}$ have been produced which output nearly $\SI{100}{\watt}$, so a conservative $\SI{80}{\watt}$ of pumping will be used in this calculation~\cite{yinHighpowerAllfiberWavelengthtunable2014}. The most successful optical cooling experiments to date have been performed in an astigmatic Herriott cell to allow for a many-pass configuration~\cite{gragossian_astigmatic_2016}. The best performing Herriott cells used for this application have given over 150 round trips, so here we will use 120 round trips with a mirror reflectivity of $1-150*10^{-6}$~\cite{gragossian_astigmatic_2016}. We additionally assume that the addition of the coolant crystal will have minimal effect on the radiative load on the system. Just like for the radiative cooling, we will assume an absorption of 1\,ppm by the coating of the PSOMA mirror. The density of Ho:BYF, used for these plots, was found in Ref.~\cite{ma_39_2021}.

\begin{figure}[ht]
\centering
\includegraphics[width=\columnwidth]{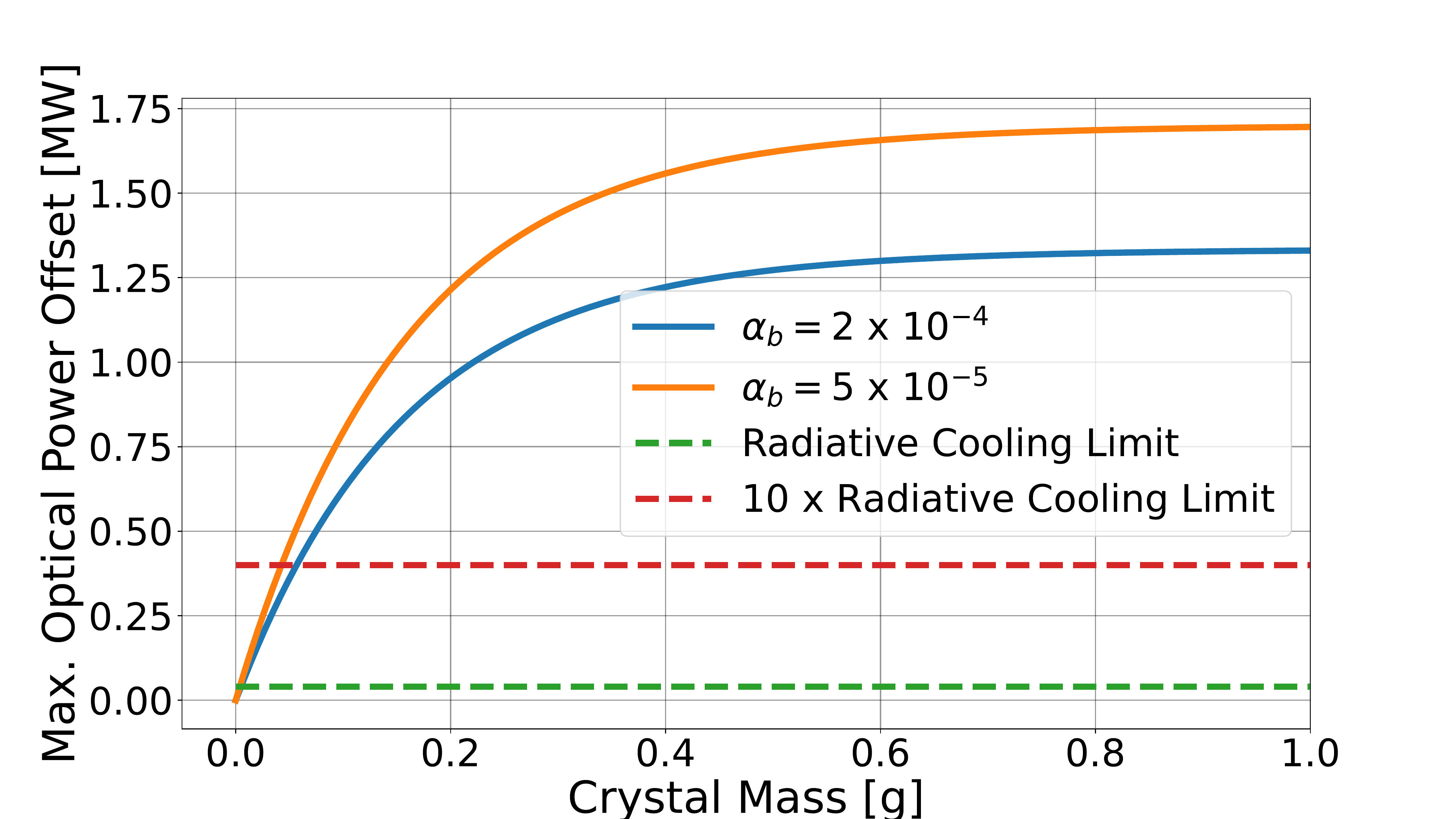} 
\caption{Plots of peak (no thermal link) cavity power offset versus crystal mass for a pair of theoretical Ho:BYF crystals at 100~K. The limit for radiative cooling is shown with a green dotted line. Ten times that limit is shown with a red dotted line. Note that existing Ho:BYF crystals are not depicted here because they do not cool at \SI{100}{\kelvin} or even 123\,K. The actual cooling power could be less than the peak if a thermal link is used.} 
\label{fig:PSOMAPc}
\end{figure}

Calculated values for offset power, defined to be the amount of additional power that could circulate in the cavity without heating the mirror when cooled by the crystal at \SI{100}{\kelvin} are plotted against mass in Figure\,\ref{fig:PSOMAPc}. Mass dependence appears because the total absorption of the crystal is limited by its volume, which is directly related to its mass.
\begin{figure}[hb!]
\begin{center}
    \includegraphics[width=\columnwidth]{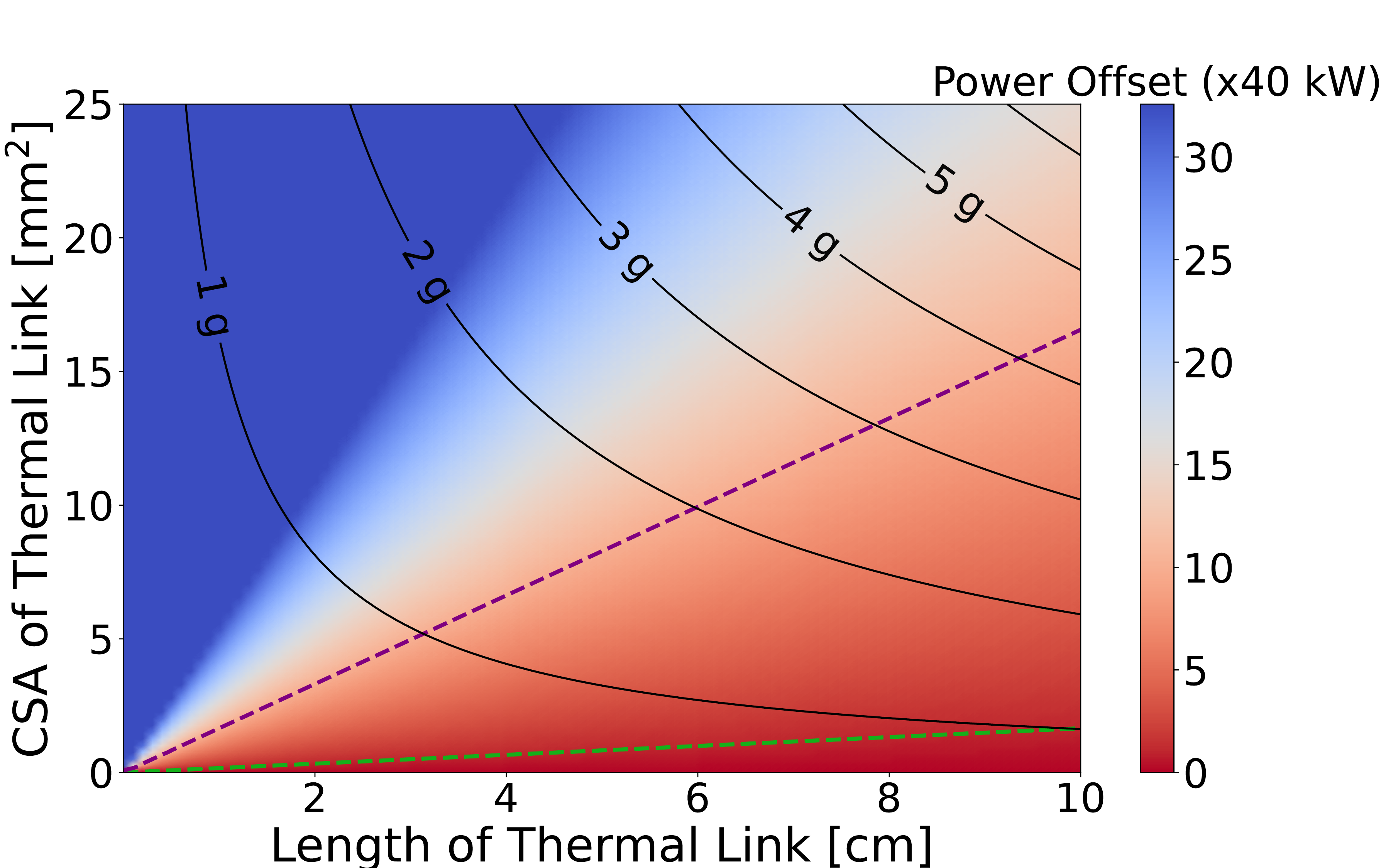}
    \caption{A 2D plot showing how the introduction of a silicon thermal link of a sampling of lengths and cross-sectional areas (CSA) affects the cooling power of an optical refrigerator. The green dotted contour shows the limit of radiative cooling power, and the purple dotted contour shows ten times that limit, which is our goal for OR here. The black contours show the combined added mass of the crystal and thermal link for the given arrangement. This plot assumes a 5~mm$^3$ Ho:BYF crystal with improved background absorption at 100~K.}
\end{center}
\label{fig:thermal_link_2d_plot}
\end{figure}

These plots show that even a very light holmium-doped crystal could match or outperform the best possible figures for cooling the mirrors only radiatively. This means that with modest and anticipated improvements in crystal properties for holmium-doped crystals, optically refrigerating the PSOMA mirrors could be viable for Voyager. In terms of per-mass cooling, even crystals improved by the smallest amount would vastly outperform radiative cooling. These plots also inform us about the most important properties to improve. Although increased doping would improve the cooling power for any crystal, the significant changes come from improvements in external quantum efficiency. Recent results for Ho:BYF crystals have shown significant improvements in external quantum efficiency as well as strong promise for high power cooling at \SI{123}{\kelvin}~\cite{rostami_observation_2021}. It is worth noting that serious research into Holmium doped coolants is still very new and some assumptions that are made here may not hold, such as the background absorption coefficient being independent of temperature. For example, for Yb:YLF, the background absorption coefficient decreases with temperature, but different mechanisms may be present in Ho-doped crystals~\cite{volpi_optical_2019}. 

\subsection{Noise Budget}
We evaluated the contribution each of the noise sources described in Sec.~\ref{sec:noiseconsider} to the readout, as well as the existing displacement noise sources of suspension thermal and cavity pump RIN for PSOMA to see if the introduction of an optical refrigerator would have a significant effect on the noise budget. The noise budget was calculated using a Finesse, a frequency domain optical simulation ~\cite{Finesse}. The optical layout of the model can be found in figure 1 of~\cite{bai2020phase}. 
The RIN amplitude spectral density of both the PSOMA and OR pumps are assumed to be
\begin{equation}
    \text{RIN}(f)=\left|\frac{f+f_0}{f}\right|\frac{1\times 10^{-9}}{\sqrt{Hz}},\quad f_0=\SI{50}{\hertz}.
\end{equation}

\begin{figure}[htb!]
\begin{center}
\includegraphics[width=\columnwidth]{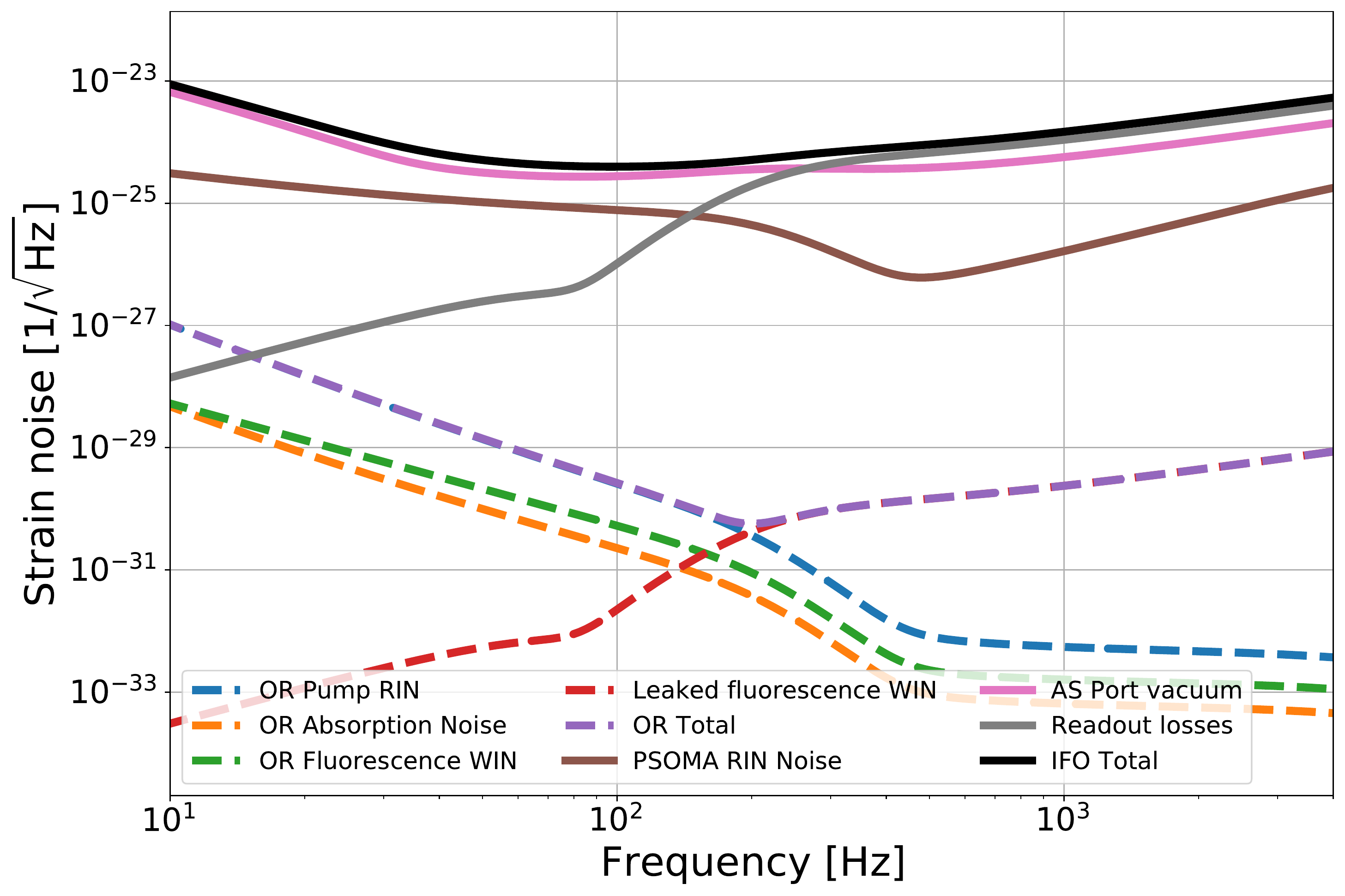}
\caption{Estimated noise curves for the noise introduced by the optical refrigerator (dashed) plotted alongside dominant noise sources in PSOMA (solid). OR absorption noise includes both absorption imbalance and fluorescence recoil noises.
PSOMA RIN is the noise coming from the RIN of the PSOMA pump laser. AS (asymmetric) Port vacuum is the quantum vacuum noise, not including noise due to loss at the readout. Readout loss is assumed to be \SI{10}{\percent}. IFO (interferometer) Total is the total noise in the Voyager+PSOMA configuration. It can be seen that the OR does not significantly impact the noise budget.
}
\label{fig:PSOMA_Noise}
\end{center}
\end{figure}

For the PSOMA RIN coupling due to radiation pressure, we assumed a circulating power of \SI{40}{\kilo\watt}. However, this figure could potentially be much higher due to the extra cooling the optical refrigerator provides. The fluorescence spectrum used for the leaked light noise is provided by S. Rostami ~\cite{rostami_low-temperature_2021} and was presented in~\cite{rostami_observation_2021}. The results for the noise budget with optical refrigeration included are shown in Figure~\ref{fig:PSOMA_Noise}. 

The noise introduced by the optical refrigerator is fairly well below the existing displacement noise in PSOMA. This is a promising sign, as this means that it is likely possible to implement optical refrigeration without a significant effect on the noise budget. As stated before, these parameters are not the optimal noise conditions for using an optical refrigerator, so it is likely that radiation pressure noise would be even less of a concern in implementation.

\section{Conclusions and Future Work}
Here we have shown that optical refrigeration (OR) may be useful in cooling mirrors or other small optical components for sensitive optomechanical experiments, such as future upgrades of LIGO at \SI{123}{\kelvin} in a way that is essentially vibration-free. In particular, we show that with upcoming improvements to coolant crystals, optical refrigeration will likely be able to outperform existing cooling methods by a factor of 10 while minimally adding to interferometer noise. We also show that a thermal link could be used to mitigate this noise further, if necessary. Beyond improvements to crystal quality, future work involves a more detailed investigation into cooler design and experimental verification of the noise curves presented here. If OR continues to show promise, it could be an important cooling method for LIGO Voyager and beyond.


\newpage

\textbf{Acknowledgments}
We want to acknowledge support from S. Rostami and M. Sheik-Bahae for providing us with the fluorescence spectrum of Ho:BYF and for the valuable discussion we had about OR. We would also like the Voyager and the advanced interferometer configurations working groups. Yehonathan thanks Gautam Venugopalan for providing preliminary noise calculations. This material is based upon work supported by NSF’s LIGO Laboratory, a major facility fully funded by the National Science Foundation.

\textbf{Disclosures} The authors declare no conflicts of interest.

\textbf{Data availability} Data underlying the results presented in this paper are not publicly available at this time but may be obtained from the authors upon reasonable request.

\bibliographystyle{unsrt85}
\bibliography{RanaGWrefs}

\appendix

\input{misc.tex}

\end{document}

%% file: misc.tex
\section{Light Transmission into the PSOMA Cavity}
\label{sec:leaked_light}
One other possible source of noise introduced by the addition of an optical refrigerator is fluorescence originating from the OR crystal leaking into the optical cavity it is attached to and traveling through to the readout. Because this light is noisy, it would contribute some noise to the readout, which can be estimated using known methods which we have adapted to fit our particular situation. We first need to estimate the coupling of the incident fluorescence to the Hermite-Gauss modes of the laser cavity.
\subsection{Coupling to Hermite-Gauss Modes}

For a two-mirror cavity, the transmitted intensity $I_t$ is given by
\begin{equation}
    I_t = \frac{T_1 T_2 I_{in}}{\left|1-\sqrt{R_1 R_2}e^{i\phi_{rt}}\right|^2},
\end{equation}
where $I_{in}$ is the incoming intensity in the HG mode, $T_{1,2},R_{1,2}$ are the transmission and reflection coefficiants of mirror number 1 and 2, respectively. These coefficients satisfy
\begin{equation}
    T_{i}+R_{i}+L_{i}=1,
\end{equation}
where $L_{i}$ is the loss in the i'th mirror which can come from absorption or scattering losses. This loss can be assumed to be on the order of $\sim$10\,ppm for the PSOMA mirrors. $\phi_{rt}$ is the roundtrip accumalated phase and is given by
\begin{equation}
    \phi_{rt}=2\pi\frac{L_{rt}}{\lambda}+(1+m+n)\arctan\left(\frac{z}{z_R}\right),
\end{equation}
where $L_{rt}$ is the round trip optical path, $\lambda$ is the wavelength of light, and $z_R$ is the Rayleigh length, determined from the geometry of the cavity. $m$ and $n$ refer to the indices of a particular Hermite-Gauss mode, where $m=n=0$ is the lowest, Gaussian, mode and $m>0$ or $n>0$ represent some higher order mode.

For high Finesse cavities, where $T,L<<R$, the peak cavity transmission can be approximated by 
\begin{equation}
    I_{peak}=I_t\left(\phi_{rt}=2\pi N\right)\approx\frac{4T_1 T_2 I_{in}}{\rho^2},
\end{equation}
where $\rho=\left(T_1+T_2+L_1+L_2\right)$ is the fraction of light lost in a round trip. In this approximation, the frequency linewidth of each mode is
\begin{equation}
    \Delta f = \frac{\text{FSR}}{\mathcal{F}}\approx\frac{c\rho}{2\pi L_{rt}},
\end{equation}
where FSR is the Free Spectral Range and $\mathcal{F}$ is the finesse of the cavity. 

The radiance of a HG mode is $\mathcal{L}=\frac{P_0}{M_x^2 M_y^2 \lambda^2}$, where $P_0$ is the total power in the mode and $M_{x,y}^2$ are the beam quality factors given by
\begin{subequations}
    \label{eq:all1}
     \begin{align}
      M_x^2=2n+1 \label{eq1} \\
      M_y^2=2m+1 \label{e2}
     \end{align}
     \end{subequations}
where n,m are the order of the HG mode.

Now we assume that the light which exits the crystal is Lambertian, which is a good approximation because. The radiance of a Lambertian source is $\mathcal{L}=\frac{P_0}{\pi A}$, where $P_0$ is the total power and A is the area of the source. Due to the conservation of Radiance or phase-space volume, the maximum coupling achievable between a Lambertian source and an Hermite Gauss mode is $\frac{M_x^2 M_y^2 \lambda^2}{\pi A}$. We will use this coefficient as an upper limit which we will call $c_{m,n}(\lambda)$.

To minimize thermal noises, LIGO cavities are designed for maximum beam sizes allowed by mirror clipping. That means, that higher order modes having larger mode size and divergence quickly become very lossy as the mode order increases. Therefore, it is a good approximation to consider only the lowest order modes up to $n=m=5$. In this approximation, the total coupling for a set of Hermite-Gauss modes, $C(\lambda)$, is given by $ C(\lambda)=\sum_{n=0}^{5}\sum_{m=0}^{5} c_{n,m}(\lambda)$.

Next, the specifications of PSOMA found in \cite{bai2020phase} tell us that the frequency bandwidth of emitted fluorescence ($\sim 8$ THz) is much larger than the free spectral range of the cavity ($\sim 10$ MHz), so we can assume that the modes are roughly evenly distributed throughout the band. Therefore, the power of fluorescence coupled to the PSOMA cavity is the number of modes multiplied by the (frequency dependent) power of light transmitted to each mode. This is given, in the form of a power spectral density, by 
\begin{equation}
    T_{HG}(f)=2*C(\frac{c}{f})*\frac{4T_1 T_2 I_{in}(f)}{\rho^2}\frac{\Delta f}{FSR}
\end{equation}
where there is a factor of two due to the presence of a pair of polarizations for each mode and $I_{in}(f)$ is the power spectral density of emission, normalized to the total power incident on the mirror. The total power transmitted is therefore just the integral of $T_{HG}(f)$ over all frequencies.

\subsection{Coupling from PSOMA to Readout}
\label{sec:readout}
Light that is leaked into the PSOMA cavity is filtered once more before it is read out by a sensor. This filter, called the Output Mode Cleaner (OMC), consists of a short bow tie cavity designed to isolate the signal mode from higher order spatial modes and modulation sidebands of light that may be travelling through the readout chain. By design, one of the modes of the OMC lines up perfectly with one of the modes of PSOMA in order to transmit that carrier mode. 

We use parameters from the current design of the OMC to estimate the effect it would have on light transmitted into the cavity. The free spectral range of the OMC is almost exactly $\frac{53}{2}$ that of PSOMA, which means that the frequencies of the two cavities will line up every 53 PSOMA modes (which is the same as two OMC modes). Note that it is important that both 53 and 2 are prime numbers, which means that there are no smaller numbers of modes for which the frequencies will line up. The bandwidth of each OMC mode is also much wider than for PSOMA, which means that when they do line up, we can assume near unity transmission for incoming PSOMA light in each mode. This means that the effect of the OMC is essentially an attenuation of the incoming optical power by a factor of 53, which reduces the incoming noise in that power by the same factor. 

Finally, the fluorescence bandwidth of Ho:BYF is a few THz and so it is, to a very good approximation, a white noise source in the GW detection band. Assuming it is spatially coherent (most pessimistic assumption) its amplitude spectral density (ASD) can be written as $\frac{I_t}{\sqrt{BW}}$, where $I_t$ is the total power of the leaked fluorescence and $BW$ is its bandwidth.

\section{Absorption of Reflected Light}
\label{sec:reabsorption}

Because the cooling crystal emits light in the wavelength band where the mirror it cools is designed to be highly reflective, one concern would be absorption of reflected light which might heat the crystal. Here we provide an upper bound on the heating due to this effect and show that it would not be a concern in implementation even in the worst case.

The effects of reflected light absorption is similar to that of reabsorption in the crystal, which has an effect on the external quantum efficiency and has been theoretically investigated \cite{heeg_influence_2005}. Borrowing from those investigations, we can estimate the average absorption cross section for reflected photons as 
\begin{equation}
    \alpha_r=\int \alpha(\lambda) F_0(\lambda) d\lambda
\end{equation}.
where $\alpha(\lambda)$ is the regular absorption coefficient and $F_0(\lambda)$ is the emission spectrum normalized such that $\int F_0 d\lambda=1$. The mean wavelength of the light absorbed in this manner is given by
\begin{equation}
     \bar{\lambda_r}= \alpha_r^{-1}\int \alpha(\lambda) F_0(\lambda)\lambda d\lambda.
\end{equation}

The process of absorbing reflected light can heat the crystal in three ways. A photon can be absorbed then re-emitted as light, absorbed by background impurities, or absorbed then not re-emitted. The total heating contribution of these processes is:
\begin{align*}
    P_{added}= &\eta_{eqe}(\frac{hc}{\bar{\lambda_r}}-\frac{hc}{\bar{\lambda_f}})\alpha_r \bar{L} P_{inc}\frac{\bar{\lambda_f}}{hc}\\
    &+\alpha_B \bar{L} P_{inc}\\
    &+(1-\eta_{eqe})\frac{hc}{\bar{\lambda_r}}\alpha_r \bar{L} P_{inc}\frac{\bar{\lambda_f}}{{hc}} \numberthis \label{eqn}
\end{align*}
where $\bar{L}$ is the average length a light ray travels through the crystal and $P_{inc}$ is the power incident on the crystal.

Due to a lack of existing absorption data for Holmium-doped crystals well below the mean fluorescence wavelength, we used an upper bound that all photons below the mean fluorescence wavelength would be absorbed and all those above the mean fluorescence wavelength would not be. We also found $P_{inc}$ by calculating the solid angle the front face of the crystal subtends on itself at a distance of twice the depth of the mirror. We then multiplied the resulting power by a factor of 3.5 to compensate for the effects of the high refractive index of silicon. We arrived at an upper bound on the total heating due to this effect of around $2$ mW, much less than the potential heat lift contributed by optical refrigeration.

\section{Excess Photon Noise}
\label{sec:WIN_noise}

Displacement noise due to photon beats---excess photon noise---requires some more careful treatment because it has not been studied before for LIGO. The statistics of broadband light was described by Hodara \cite{hodara_statistics_1965}. We apply those statistics to ascertain the noise contribution of the broadband fluorescence which reflects off the front surface of the PSOMA mirror. 

We start by breaking the PSOMA mirror surface area $A$ into $N$ coherence areas $A_c$ where
\begin{equation}
	A=NA_c.
\end{equation}
At each coherence area the radiation pressure force is
\begin{equation}
	F_c\left(t\right)=\frac{A_c}{c}I_c\left(t\right),
\end{equation}
where $I_c$ is the optical energy flux density incident on that coherence area and $c$ is the speed of light in vacuum. The total force acting on the PSOMA mirror is
\begin{equation}
	F\left(t\right)=\sum_{k=1}^{N}F_{ck}\left(t\right)=\frac{A_c}{c}\sum_{k=1}^{N}I_{ck}\left(t\right).
\end{equation}
The autocorrelation function of the force function is then given by
\begin{equation}
	\left<F\left(t\right)F\left(t+\tau\right)\right>=\left(\frac{A_c}{c}\right)^2\sum_{k=1}^{N}\sum_{j=1}^{N}\left<I_{ck}\left(t\right)I_{cj}\left(t+\tau\right)\right>,
\end{equation}
where angular brackets denote ensemble average. Since the intensity at different coherence areas is by definition uncorrelated, all the cross terms vanish. Additionally, assuming spatial homogeneity we get
    \begin{align}
		\left<F\left(t\right)F\left(t+\tau\right)\right>&=\left(\frac{A_c}{c}\right)^2N\left<I_{c}\left(t\right)I_{c}\left(t+\tau\right)\right>\\
		&=\frac{A_cA}{c^2}\left<I_{c}\left(t\right)I_{c}\left(t+\tau\right)\right>.
    \end{align}
The force ASD being the root of the Fourier transform of the autocorrelation function is given by
\begin{equation}
	A_{FF}\left(f\right) =\frac{\sqrt{A_cA}}{c}A_{II}\left(f\right),
\end{equation}
where $A_{II}$ is the optical intensity ASD. Assuming that the ASD is constant over the entire emission spectrum we can write $A_{FF}$ in terms of the fluorescence total optical power $I_{tot}$ as
\begin{equation}
	A_{FF}\left(f\right)  =\frac{1}{c}\sqrt{\frac{A_c}{A}}\frac{I_{tot}}{\sqrt{BW}},
\end{equation}
where $\sqrt{BW}$ is the bandwidth of the fluorescence spectrum. Assuming the PSOMA mirror is a free mass, the displacement noise due to radiation pressure is given by
\begin{equation}
	A_{xx}\left(f\right) = \frac{1}{4\pi^2mcf^2}\sqrt{\frac{A_c}{A}}\frac{I_{tot}}{\sqrt{BW}},
\end{equation}
where $m$ is the mass of the PSOMA mirror. We then use the displacement to strain transfer function to evaluate the effect of this noise on the total strain noise budget. 

The average coherence area is estimated using the an approximation given in Eq. 2.4 of Ref. \cite{mandel1965coherence}. This approximation is given by $A_c \sim \alpha (\frac{\lambda R}{\Delta l})^2$, where $\lambda$ is the mean fluorescence wavelength, $R$ is the distance from source to detector, $w$ is a defining width or radius of the source, and $\alpha \sim 1$ is a prefactor of order unity related to the geometry of the source. The approximation is meant to apply to the far field, which strictly does not describe the relationship between the mirror surface and the coolant crystal. However, because coherence area increases monotonically with distance from the source and is 0 at the source, this approximation will be an overestimate in the near field, giving us an upper limit on the WIN. Our estimated values for these parameters are shown in Table . This fluctuation in the photon arrival count is then used to find a the magnitude of the force, which is converted to a displacement noise magnitude.

%% file: forArxiv.bbl
\begin{thebibliography}{10}

\bibitem{GW150914}
{LIGO Scientific Collaboration and Virgo Collaboration}.
\newblock Observation of gravitational waves from a binary black hole merger.
\newblock {\em Phys. Rev. Lett.}, 116:061102, Feb 2016.

\bibitem{theligoscientificcollaborationAdvancedLIGO2015}
{The LIGO Scientific Collaboration}, J.~Aasi, B.~P. Abbott, et~al.
\newblock Advanced {{LIGO}}.
\newblock {\em Classical and Quantum Gravity}, 32(7):074001, April 2015.

\bibitem{Acernese_2014}
F.~Acernese, M.~Agathos, K.~Agatsuma, et~al.
\newblock Advanced virgo: a second-generation interferometric gravitational
  wave detector.
\newblock {\em Classical and Quantum Gravity}, 32(2):024001, dec 2014.

\bibitem{gwtc-3}
{The LIGO Scientific Collaboration}, {The Virgo Collaboration}, {The KAGRA
  Collaboration}, et~al.
\newblock Gwtc-3: Compact binary coalescences observed by ligo and virgo during
  the second part of the third observing run, 2021.

\bibitem{VoyagerInst:2020}
R.~X. {Adhikari}, K.~{Arai}, A.~F. {Brooks}, et~al.
\newblock {A cryogenic silicon interferometer for gravitational-wave
  detection}.
\newblock {\em Classical and Quantum Gravity}, 37(16):165003, August 2020.

\bibitem{reitzeCosmicExplorerContribution2019}
D.~Reitze, R.~X. Adhikari, S.~Ballmer, et~al.
\newblock Cosmic {{Explorer}}: {{The U}}.{{S}}. {{Contribution}} to
  {{Gravitational-Wave Astronomy}} beyond {{LIGO}}.
\newblock {\em Bulletin of the AAS}, 51(7), 9 2019-09-30.

\bibitem{O3:Squeezing:2019}
M.~Tse, H.~Yu, N.~Kijbunchoo, et~al.
\newblock Quantum-enhanced advanced ligo detectors in the era of
  gravitational-wave astronomy.
\newblock {\em Phys. Rev. Lett.}, 123:231107, Dec 2019.

\bibitem{miao2019quantum}
H.~Miao, N.~D. Smith, and M.~Evans.
\newblock Quantum limit for laser interferometric gravitational-wave detectors
  from optical dissipation.
\newblock {\em Physical Review X}, 9(1):011053, 2019.

\bibitem{bai2020phase}
Y.~Bai, G.~Venugopalan, K.~Kuns, et~al.
\newblock Phase-sensitive optomechanical amplifier for quantum noise reduction
  in laser interferometers.
\newblock {\em Physical Review A}, 102(2):023507, 2020.

\bibitem{shapiro_cryogenically_2017}
B.~Shapiro, R.~X. Adhikari, O.~Aguiar, et~al.
\newblock Cryogenically cooled ultra low vibration silicon mirrors for
  gravitational wave observatories.
\newblock {\em Cryogenics}, 81:83--92, January 2017.

\bibitem{seletskiy_laser_2016}
D.~V. Seletskiy, R.~Epstein, and M.~Sheik-Bahae.
\newblock Laser cooling in solids: advances and prospects.
\newblock {\em Reports on Progress in Physics}, 79(9):096401, September 2016.

\bibitem{rostami_observation_2019}
S.~Rostami, A.~R. Albrecht, A.~Volpi, and M.~Sheik-Bahae.
\newblock Observation of optical refrigeration in a holmium-doped crystal.
\newblock {\em Photonics Research}, 7(4):445, April 2019.

\bibitem{rostami_observation_2021}
S.~Rostami and M.~Sheik-Bahae.
\newblock {Observation of optical refrigeration in Ho:BYF crystal}.
\newblock In D.~V. Seletskiy, M.~Sheik-Bahae, and M.~K. Kuno, editors, {\em
  Photonic Heat Engines: Science and Applications III}, volume 11702.
  International Society for Optics and Photonics, SPIE, 2021.

\bibitem{hehlen_first_2018}
M.~P. Hehlen, J.~Meng, A.~R. Albrecht, et~al.
\newblock First demonstration of an all-solid-state optical cryocooler.
\newblock {\em Light Sci Appl}, 7(1):15, June 2018.
\newblock Number: 1 Publisher: Nature Publishing Group.

\bibitem{yu_quantum_2020}
H.~Yu, L.~McCuller, M.~Tse, et~al.
\newblock Quantum correlations between light and the kilogram-mass mirrors of
  {LIGO}.
\newblock {\em Nature}, 583(7814):43--47, July 2020.

\bibitem{wang_boosting_2022}
C.~Wang, C.~Zhao, X.~Li, et~al.
\newblock Boosting the sensitivity of high-frequency gravitational wave
  detectors using \${PT}\$-symmetry.
\newblock {\em Physical Review D}, 106(8):082002, October 2022.
\newblock Publisher: American Physical Society.

\bibitem{miao_enhancing_2015}
H.~Miao, Y.~Ma, C.~Zhao, and Y.~Chen.
\newblock Enhancing the {Bandwidth} of {Gravitational}-{Wave} {Detectors} with
  {Unstable} {Optomechanical} {Filters}.
\newblock {\em Physical Review Letters}, 115(21):211104, November 2015.
\newblock Publisher: American Physical Society.

\bibitem{cripe_measurement_2019}
J.~Cripe, N.~Aggarwal, R.~Lanza, et~al.
\newblock Measurement of quantum back action in the audio band at room
  temperature.
\newblock {\em Nature}, 568(7752):364--367, April 2019.
\newblock Number: 7752 Publisher: Nature Publishing Group.

\bibitem{hodara_statistics_1965}
H.~Hodara.
\newblock Statistics of thermal and laser radiation.
\newblock {\em Proceedings of the IEEE}, 53(7):696--704, 1965.

\bibitem{mandel1965coherence}
L.~Mandel and E.~Wolf.
\newblock Coherence properties of optical fields.
\newblock {\em Reviews of Modern Physics}, 37(2):231--287, April 1965.

\bibitem{Kip:Scatter95}
E.~E. Flanagan and K.~S. Thorne.
\newblock Scattered-light noise for ligo.
\newblock Technical Report T950102, LIGO, 1995.

\bibitem{rostami_low-temperature_2021}
S.~Rostami.
\newblock Private communication, 2021.

\bibitem{sheik-bahae_role_2022}
M.~Sheik-Bahae and J.~Kock.
\newblock The role of absorption saturation and amplified spontaneous emission
  in cryogenic optical refrigeration.
\newblock In {\em Photonic {Heat} {Engines}: {Science} and {Applications}
  {IV}}, volume 12018, pages 9--19. SPIE, March 2022.

\bibitem{yinHighpowerAllfiberWavelengthtunable2014}
K.~Yin, B.~Zhang, G.~Xue, L.~Li, and J.~Hou.
\newblock High-power all-fiber wavelength-tunable thulium doped fiber laser at
  2 {$M$}m.
\newblock {\em Optics Express}, 22(17):19947, August 2014.

\bibitem{gragossian_astigmatic_2016}
A.~Gragossian, J.~Meng, M.~Ghasemkhani, A.~R. Albrecht, and M.~Sheik-Bahae.
\newblock Astigmatic {Herriott} cell for optical refrigeration.
\newblock {\em Optical Engineering}, 56(1):011110, December 2016.

\bibitem{ma_39_2021}
C.~Ma, Y.~Zhang, J.~Guo, et~al.
\newblock A 3.9 µm {Ho3}+:{BaY2F8} laser directly pumped by laser diodes.
\newblock {\em Electronics Letters}, 57(20):779--781, 2021.
\newblock \_eprint: https://onlinelibrary.wiley.com/doi/pdf/10.1049/ell2.12250.

\bibitem{volpi_optical_2019}
A.~Volpi, J.~Meng, A.~Gragossian, et~al.
\newblock Optical refrigeration: the role of parasitic absorption at cryogenic
  temperatures.
\newblock {\em Optics Express}, 27(21):29710, October 2019.

\bibitem{Finesse}
D.~D. Brown and A.~Freise.
\newblock Finesse, May 2014.
\newblock {You can download the binaries and source code at
  \url{http://www.gwoptics.org/finesse}.}

\bibitem{heeg_influence_2005}
B.~Heeg, P.~A. DeBarber, and G.~Rumbles.
\newblock Influence of fluorescence reabsorption and trapping on solid-state
  optical cooling.
\newblock {\em Applied Optics}, 44(15):3117--3124, May 2005.

\end{thebibliography}
